\documentclass[aps,prl,superscriptaddress,twocolumn
]{revtex4-1}
\usepackage{graphicx}
\usepackage{amsmath}
\usepackage{mathtools}
\usepackage{amsfonts}
\usepackage{bm}
\usepackage[normalem]{ulem}
\usepackage[usenames, dvipsnames]{xcolor}

\usepackage{hyperref}
\hypersetup{colorlinks=true, linkcolor=blue, citecolor=blue, urlcolor=blue}

\begin{document}

\title{Magnetic penetration depth and $T_c$ in superconducting nickelates}

\author{F. Bernardini}
\affiliation{Dipartimento di Fisica, Universit\`a di Cagliari, IT-09042 Monserrato, Italy}

\author{V. Olevano}
\affiliation{Institut N\'eel, CNRS \& UGA, 38042 Grenoble, France}

\author{A. Cano}
\affiliation{Institut N\'eel, CNRS \& UGA, 38042 Grenoble, France}
\date{\today}

\begin{abstract}
We compute the nominal magnetic penetration depth of $R$NiO$_2$ ($R =$ La, Nd) from first principles calculations and discuss the results in relation to the superconducting $T_c$. 
We find a marked discrepancy with the well established phenomenology that correlates these two quantities in cuprates (Uemura plot). We also consider the 2D ultrathin limit and estimate the maximum attainable $T_c$ to be $\sim 180$~K according to the Nelson-Kosterliz universal relation between the superfluid density and the transition temperature.  
\end{abstract}

\maketitle

The recent discovery of superconductivity in Sr-doped NdNiO$_2$/SrTiO$_3$ thin films \cite{li-nature19} 
has attracted an instantaneous research attention (see e.g. \cite{botana19,wu19,sakakibara19,nomura19,hepting19,jiang19,zhou19,hirsch19,zhang19}). After many attempts, this can potentially be the first successful extension of high-$T_c$ cuprate superconductivity to isostructural/isoelectronic nickelates. This long sought extension is expected to shed light on their microscopic superconducting mechanism, even if it poses some important challenges to the current paradigm. In particular, the parent nickelates are metals without magnetic order as opposed to their cuprate counterparts that are antiferromagnetic charge-transfer-insulators. Consequently, Cooper pairing in (Nd,Sr)NiO$_2$ seems to result from a rather different normal non-superconducting state. 

First-principles calculations based on density-functional-theory (DFT) consistently find that, among the five Ni 3$d$ bands, only the 3$d_{x^2-y^2}$ one intersects the Fermi level \cite{pickett-prb04,botana19,wu19,sakakibara19,nomura19,hepting19}. This establishes a promising analogy to CaCuO$_2$, i.e. the parent compound of high-$T_c$ cuprates. In fact, the electron-phonon coupling has been ruled out as the exclusive origin of the observed superconductivity in (Nd,Sr)NiO$_2$ \cite{nomura19}. At the same time, electrons in the Nd layer make additional electron pockets in the Fermi surface that likely prevent the system from being a Mott insulator, with Kondo physics potentially playing a role \cite{hepting19,zhang19}. Besides, the charge transfer gap between the Ni 3$d$ states and O 2$p$ states is larger than that in cuprates \cite{jiang19}. On the other hand, spin fluctuations may still be important for superconductivity, even if there is no long-range magnetic order. In that case, the dominant pairing has been proposed to yield a $d$-wave superconducting gap \cite{wu19,sakakibara19} with a distinct spin resonance feature that can be tested experimentally \cite{zhou19}. Here, we compute the nominal magnetic penetration depth of the superconducting nickelates as a function of hole doping and discuss these theoretical results in relation to the observed $T_c$. 

The magnetic penetration depth $\lambda$ is a fundamental quantity of superconductors as it is pivotal to explain the Meissner effect \cite{tinkham2004}. This quantity can be determined experimentally by means of different complementary techniques such as the tunnel diode oscillator technique \cite{Prozorov_2006,Prozorov_2011} and muon-spin resonance ($\mu$SR)  \cite{uemura-prl89,amato,sonier-rmp00,Bhattacharyya-18}. The temperature dependence of $\lambda$ maps the amount of excited quasiparticles and thereby the structure of the superconducting gap. However, in the London approximation, the zero-temperature magnetic penetration depth in the clean limit reduces to $\lambda_L(0) = \sqrt{ {m^* \over  \mu_0 n_s e^{*2} }}$, where $m^*$ is the effective mass of charge carriers, $\mu_0$ is the vacuum permeability, $n_s$ is the charge carrier density, and $e^*$ is an effective electron charge. 
Thus, it basically becomes a band-structure property formally unrelated to the gap function. In fact, following Eilenberger's formulation of superconductivity \cite{chandrasekhar93,kogan-prb02}, a band-structure-specific result can be obtained as
\begin{align}    
(\lambda^{2})_{ij}^{-1}(0)= \frac{\mu_0e^2}{4\pi^3\hbar}\oint_\text{FS} dS\frac{{ v}_{Fi}{ v}_{Fj}}{v_F},
\label{lambda}
\end{align}
where the integral is over the Fermi surface with ${\bf v}_F$ being the Fermi velocity (which is a measure of the DOS).

\begin{figure*}[t!]
\includegraphics[scale=0.6]{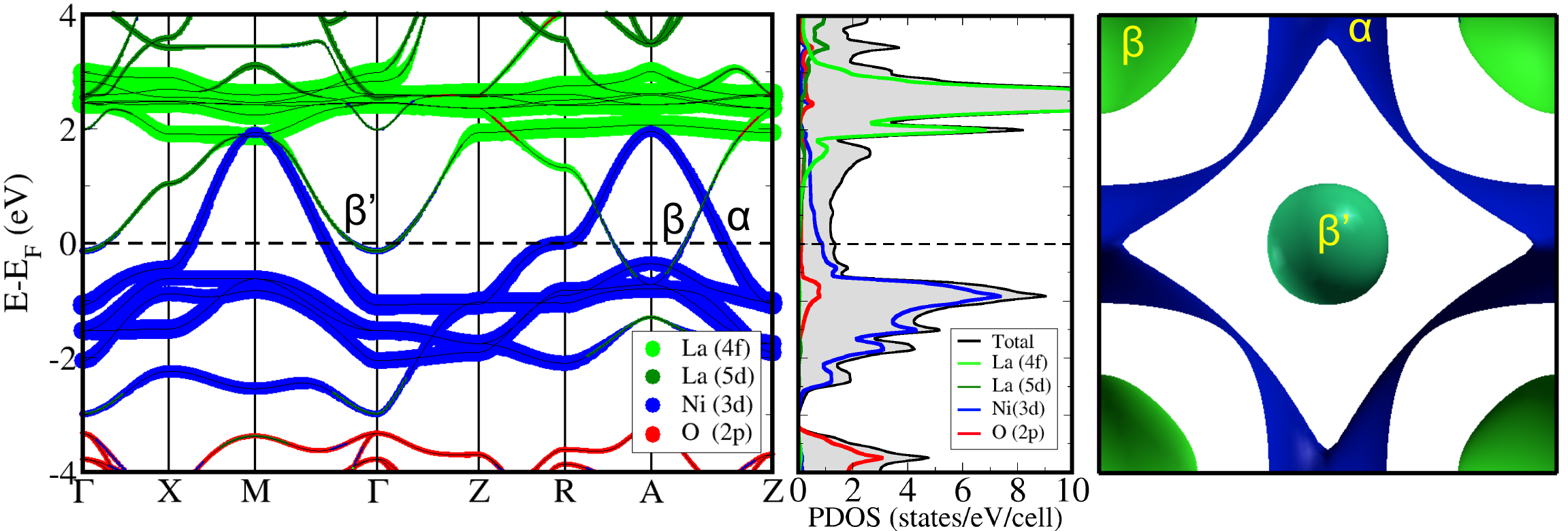}
  \caption{First-principles band structure  (left, ``fatband" plot), orbital resolved density of states (middle) and top view of Fermi surface (right) of LaNiO$_2$.}  
\label{bands}
\end{figure*}

We computed Eq. \eqref{lambda} from DFT calculations that conveniently reproduce the reported band structure of the La and Nd nickelates. 
Specifically, we used the FLAPW method as implemented in the {\sc{WIEN2k}} package \cite{Wien2k} with the LDA exchange-correlation functional \cite{LDA}. 
In order to avoid the ambiguous treatment of the $f$-orbital bands, we followed \cite{botana19,sakakibara19} and focused on the mother compound LaNiO$_2$. Further, we modeled Sr doping as a rigid shift of the Fermi level as in \cite{wu19}. We also considered NdNiO$_2$ with the Nd-4$f$-states in the core as well as the influence of epitaxial strain, with which we obtained almost identical results. We performed spinless calculations with muffin-tin radii of 2.5 a.u., 2.1 a.u., and 1.62 a.u. for the La (Nd), Ni, and O atoms respectively and a plane-wave cutoff $R_{\rm mt}K_{\rm max}=7.0$. 
The integration over the Brillouin zone was performed using a 11$\times$11$\times$14 $k$-mesh for the self-consistent calculations, while a dense 48$\times$48$\times$48 $k$-mesh was used to compute and study the Fermi surface. In our calculations the Fermi velocity is directly obtained from the expectation value of the momentum operator ${\bf p}$ (${\bf v}_F = {\bf p}_F/m$), and the dense $k$-mesh was used to further perform the Fermi-surface integral. 
The reference band structure obtained in this way is illustrated in Fig. \ref{bands}. 

\begin{table}[b!]
 \begin  {tabular}{c c c c c c c }
\hline \hline    
& doping &  FS  &  $\lambda_x(0)$ [nm] & $\lambda_z(0)$ [nm] &  $\lambda_{\rm eff}(0)$ [nm]\\ \hline
LaNiO$_2$ & 0 & $\alpha$  &   47  & 245 & 61\\
& & $\beta$     & 140  & 120 & 133 \\
& & $\beta'$    &   365  & 300  & 340 \\
& & total       &   44  & 101 & 54 \\ \cline{2-6}
& 0.1 & $\alpha$  & 49  & 170  & 62  \\
&  & $\beta$      & 165 & 140 & 156 \\
& & total         & 47 & 108 & 57 \\ \cline{2-6}
& 0.2 & $\alpha$  &   50  & 215 & 64 \\
& & $\beta$  & 205  & 175 &  194 \\
& & total & 48 & 136 & 60 \\ \cline{2-6}
& 0.3 & $\alpha$  &   51  & 160 & 64 \\
& & $\beta$  & 275  & 240 &  262 \\
& & total & 50 & 133 & 62 \\ \cline{2-6}
& 0.4 & $\alpha$  &   54  & 163 & 68 \\
& & $\beta$  & 450  & 580 &  486 \\
& & total & 54 & 156 & 67 \\ \cline{1-6}
NdNiO$_2$ & 0.2 & $\alpha$  &   49  & 180 & 62 \\
          & & $\beta$  & 190  & 160 &  179 \\
          & & $\beta'$ & 290 & 220 &  262\\
& & total & 47 & 105 & 57 \\
\hline \hline
\end{tabular}
\caption{Zero-temperature magnetic penetration length obtained from DFT calculations in the London approximation for different values of Sr (hole) doping (modeled as a rigid shift of the Fermi level). The effective lambda is defined as $\lambda_{\rm eff} = 3^{1/4}[1+2 (\lambda_x/\lambda_z)^2 ]^{-1/4}\lambda_{x}$, as probed by $\mu$SR in policrystalline samples \cite{fesenko1991}. The values for NdNiO$_2$ are obtained assuming the Nd-4$f$-states in the core. }
\label{LondonDFT}
\end{table}

The resulting values for the nominal $\lambda(0)$ are collected in Table \ref{LondonDFT}. They confirm that (Nd,Sr)NiO$_2$ is a type-II superconductor (i.e. $\kappa \equiv \lambda /\xi > 1/\sqrt{2}$, with $\xi = 3.25$~nm being the Ginzburg-Landau coherence length \cite{li-nature19}). In the case of the overall in-plane component $\lambda_x(0)$, the main contribution originates from the main hole pocket $\alpha$ and does not vary dramatically with doping. The out-of-plane component $\lambda_z(0)$, in contrast, is initially dominated by the electron pocket $\beta$ and therefore undergoes a more substantial change as $\beta$ shrinks with doping. Note that, despite the apparent 2D character of $\alpha$ \cite{pickett-prb04}, the anisotropy of this contribution is moderate compared to that in cuprates and displays a non-monotonous behavior with doping. Beyond that, the effective $\lambda_{\rm eff}$ probed by $\mu$SR turns out to be essentially that of the hole pocket $\alpha$ (with $\beta$ having the effect of reducing the anisotropy resulting from $\alpha$ only).

\begin{figure}[t!]
\includegraphics[scale=0.6]{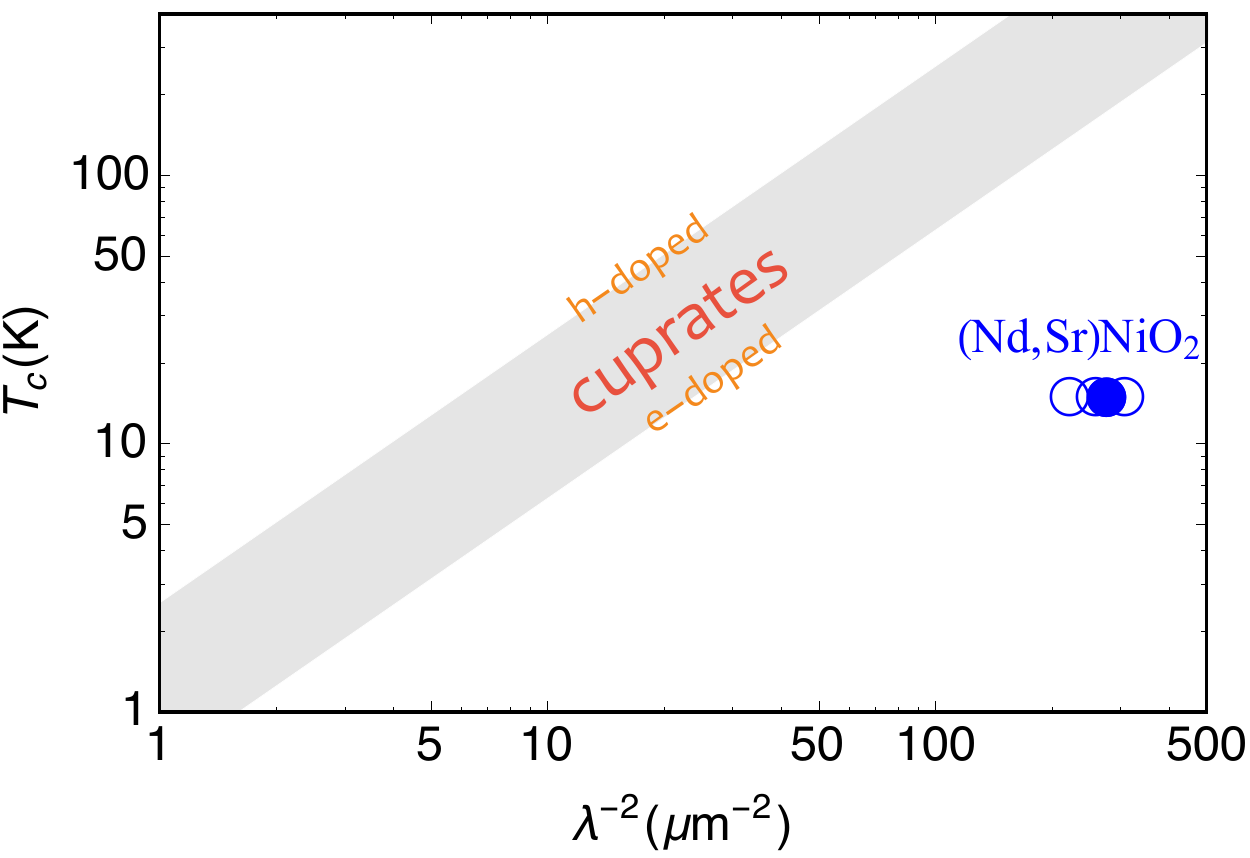}
  \caption{Uemura log-log plot for the zero-temperature effective magnetic penetration depth of (Nd,Sr)NiO$_2$ computed from Eq. \eqref{lambda}. The solid circle corresponds to 20\% Sr doping and $T_c = 15$~K, while the open circles are associated to the 10\%, 30\% and 40\% dopings in Table \ref{LondonDFT} assuming the same $T_c$.}  
\label{uemura-plot}
\end{figure}

These results are summarized in Fig. \ref{uemura-plot} in a Uemura plot. As we can see, the calculated $\lambda_{\rm eff}$'s are totally off the expected values for a cuprate with $T_c \approx 15$~K (shaded region). We note that a presumably related discrepancy has been pointed out for the Hall resistance of the parent compound \cite{botana19}. The modification of the effective mass can to some extent improve the agreement in the Uemura plot. However, the mass has to be $>10$ times larger to do the job. We note also that cuprate trend is regained if only the electron pocket $\beta$ is considered at 20\% Sr doping, although one would tend to think that this is rather fortuitous coincidence.  
Thus, reconciling these results with the well established phenomenology of cuprates seems to require a rather significant modification of the corresponding band structure (this, or the system is to be understood as a failed room-temperature superconductor according to its $\lambda$). 

Alternatively, superconductivity in (Nd,Sr)NiO$_2$ thin films can be a 2D phenomenon and hence a different rationale must be applied. This will be naturally the case in the ultrathin limit, and will also be relevant if superconductivity is driven by the interface with the substrate. The magnetic penetration depth, being a measure of the superfluid density, is also related to the superfluid stiffness $D_s$. This relation can be exploited to set bounds on the superconducting transition temperature since fluctuations of the phase of the superconducting order parameter will be the ultimate limiting factor in 2D \cite{randeira-prx19}. Such a bound directly reads from Nelson-Kosterlitz universal jump of the superfluid density \cite{nelson-prl77}:  
\begin{align}    
k_B T_c \leq \pi D_s/2.
\label{Tc}\end{align}

The 2D superfluid stifness can be estimated from our previous calculations as $D_s \approx  {\hbar^2 \over 4\mu_0 e^2}
{d \over 2\pi } \lambda^{-2}_{x}$, where $d$ is the interlayer spacing (i.e. the $c$ lattice parameter). This gives a maxium $T_c$ of about $145$~K. By restricting the integral \eqref{lambda} to the $k_{Fz}=0$ line of the 3D Fermi surface this value increases to $180$~K. While formally rigorous, this estimate has to be understood as a rather conservative upper bound since the superfluid density at $T=0$, and hence the corresponding stiffness, can reasonably be assumed to overestimate that at $T_c$.    

In summary, we have computed the zero-temperature magnetic penetration depth $\lambda (0)$ of the newly superconducting nickelate NdNiO$_2$ relying on first-principles DFT calculations to fully take into account its band-structure specific features. $\lambda (0)$ is a fundamental descriptor of superconductivity displaying a phenomenological correlation to $T_c$ in cuprates and in other unconventional superconductors. 
Our calculations confirm the system as a type-II superconductor. The in-plane component of $\lambda (0)$ is found to be dominated by the hole Fermi-surface pocket and no substantial change is obtained with doping. However, the extra electron pocket has a non-negligible impact on the eventual anisotropy. 
Remarkably, the nominal $\lambda (0)$ and the reported $T_c$ do not follow the same correlation observed in the cuprates. If the same correlation were to apply, NdNiO$_2$ would be a room-temperature superconductor. This suggests that either the reported band structure needs to be revisited  or the superconducting nature of nickelates is different.
In the 2D case relevant for the ultrathin limit, the maximum attainable $T_c$ is estimated to be $\sim 180$~K from the Nelson-Kosterlitz universal jump of the superfluid density.      

\bibliography{bib.bib}

\end{document}